\documentclass[12pt]{iopart}
\usepackage{epsfig}
\usepackage{amssymb}

\begin{document}

\title[Common Underlying Dynamics in an Emerging Market]{Common Underlying Dynamics in an Emerging Market: From Minutes to Months }

\author{Renato Vicente\dag\ \footnote[3]{To
whom correspondence should be addressed (rvicente@if.usp.br)}, Charles M. de Toledo\ddag\ , Vitor B.P. Leite \P\ and Nestor Caticha\dag\ 
}

\address{\dag\ Dep. de F{\'\i}sica Geral, Instituto de F{\'\i}sica, Universidade de S\~ao Paulo, Caixa Postal 66318, 05315-970 S\~ao Paulo - SP, Brazil}
\address{\ddag\ BOVESPA - S\~ao Paulo Stock Exchange, R. XV de Novembro, 275, 01013-001 S\~ao Paulo - SP, Brazil}
\address{\P\ Dep. de F{\'\i}sica, IBILCE, Universidade Estadual Paulista, 15054-000 S\~ao Jos\'e do Rio Preto - SP, Brazil}

\begin{abstract}
We analyse a period spanning 35 years of activity in the S\~ao Paulo Stock Exchange Index (IBOVESPA) and show that the Heston model with stochastic volatility is capable of explaining price fluctuations for time scales ranging from 5 minutes to 100 days with a single set of parameters.  We  also show that the Heston model is inconsistent with the observed behavior of the volatility autocorrelation function. We deal with the latter inconsistency by introducing a slow  time scale to the model. The fact that the price dynamics in a period of  35 years of macroeconomical  unrest may be modeled by the same stochastic process is evidence for a general underlying microscopic market dynamics.   
 \end{abstract}
\pacs{02.50.-r,89.65.-s}
\maketitle

\section{Introduction}
\label{intro}
In the last decades much attention has been devoted to the modeling of asset  returns by the Finance  community, the major drive behind this  effort being the improvement of pricing techniques for derivative contracts \cite{fouque}.  The pricing problem is  amenable to analytical solution for some stochastic volatility models \cite{gysels} such as Hull-White \cite{hullwhite}, Stein-Stein \cite{steinstein} and Heston  \cite{heston}.  Despite differences in methods and emphasis,  the cross fecundation between Economics and Physics, which dates back to the early nineteenth century (see \cite{roehner} and \cite{mirowski}), has intensified recently \cite{stanley}. Following the tradition of statistical physics, substantial effort has been made to find models capable of elucidating the basic mechanisms behind recurrent features of financial time series such as: returns aggregation (probability distributions at any time scale) \cite{yakovenko,silva}, volatility clustering \cite{engle}, leverage effect (correlation between returns and volatilities) \cite{bouchaudprl,perello}, conditional correlations \cite{boguna,lebaron}.     

Recently, a semi-analytical solution for the Fokker-Planck equation describing the distribution of log-returns  in the Heston model has been proposed \cite{yakovenko}. The authors were able to show a good agreement between the return distributions of a number of developed market stock indices and the model, for time scales spanning a wide interval ranging from 1 to 100 days.

In this  paper we show strong evidence that the Heston model is capable of explaining the return distribution of the  S\~ao Paulo Stock Exchange Index (IBOVESPA) in a period that span 35 years  of political and economical unrest, with  hyperinflation periods, currency crises and major regulatory changes. We also show that the Heston model can explain the diffusive process of IBOVESPA from minutes to months with a single set o parameters. However, we observe that the Heston model is inconsistent with the measured volatility autocorrelation function and we introduce an extension to the model along the lines discussed in \cite{bouchaud} which exhibits the correct behavior.    

\section{The Heston Model}
The Heston Model describes the dynamics of stock prices $S_t$ as a geometric Brownian motion with volatility given by a Cox-Ingersoll-Ross (or Feller) mean-reverting dynamics. In the It\^o differential form the model reads: 
\begin{eqnarray}
\label{eq_heston}
dS_t &=& S_t\,\mu_t dt\;+\;S_t\sqrt{v_t}\;dW_0(t)\\
dv_t &=& -\gamma\left[v_t-\theta\right]dt\;+\;\kappa\sqrt{v_t}\;dW_1(t),\nonumber 
\end{eqnarray}
where $v_t$ is the volatility and $dW_j$ are Wiener processes with 
\begin{equation}
\label{eq_Wiener}
\langle dW_j(t) \rangle = 0,\;\;\;\;\;\;\langle dW_j(t)\, dW_k(\tilde{t}) \rangle =  \left[\delta_{jk}+(1-\delta_{jk})\rho\right]\, \delta(t-\tilde{t})\,dt.
\end{equation}
The term $\sqrt{v_t}$ avoids unphysical negative volatilities, $\theta$ is the macroeconomic long term volatility and  $\mu_t$ represents a drift also at macroeconomic scales.  

Interestingly, $v(t)\equiv\sum_{j=1}^{d} X_j^2(t)$ is a Feller process   (\ref{eq_heston}) if $X_j$ are Ornstein-Uhlenbeck processes (OU) defined as \cite{shreve}: 
\begin{equation}
\label{eq_OU}
dX_j(t)=-\frac{b}{2}\, X_j(t)\, dt + \frac{a}{2}\, dW_j(t),
\end{equation}
where $dW_j$ describe $d$ independent Wiener processes. This observation restricts the possible parameter values to $d\ge2$ as the probability density of the volatility must vanish at zero to be consistent with the empirical data. To see how the volatility process in (\ref{eq_heston}) emerges from OU processes we apply It\^o's Lemma to get:
\begin{equation}
\label{eq_volprocess_2}
dv_t=-b\;dt\sum_{j}^{d} X_j^2 +a\sum_{j}^{d}X_j\;dW_j+\frac{a^2}{4}\;\sum_{j}^{d}dW_j^2.
\end{equation}
Using the definition of $v$ and the properties of the Wiener processes it follows that:
\begin{equation}
\label{eq_volprocess}
dv_t=\left[\frac{d}{4} \;a^2-b v_t\right]\;dt \;+\;a\sqrt{v_t}\;dW.
\end{equation}
The volatility process in (\ref{eq_heston}) can be recovered by a few variable choices: $a=\kappa$,
 $b=\gamma$ and $\theta=\frac{d}{4}\frac{\kappa^2}{\gamma}$. Note that, given the dimension $d$, there are only two free parameters in (\ref{eq_heston}). The OU processes in (\ref{eq_OU}) may be regarded as the primary microscopic sources of volatility and the condition of  non-vanishing volatility implies that $\alpha\equiv\frac{2\gamma \theta}{\kappa^2}>1$.

\section{The Fokker-Planck Equation Solution}
We now outline the solution of the Fokker-Planck equation (FP) describing the distribution of log-returns proposed by Dr$\breve{a}$gulescu and Yakovenko \cite{yakovenko}.

As we are mainly concerned with price fluctuations, we simplify equation (\ref{eq_heston}) by introducing log-returns in a window $t$ as $r(t)=\ln(S(t))-\ln(S(0))$. Using Ito's lemma and changing variables by making $ x(t)=r(t)-\int_0^t\,d\tau\mu_\tau$ we  obtain a detrended version of the return dynamics that reads:
\begin{equation}
\label{eq_detrended}
dx=-\frac{v_t}{2}\,dt+\sqrt{v_t}\,dW_0.
\end{equation}

The FP equation is, therefore, given by:
\begin{eqnarray}
\label{eqFKequation}
\frac{\partial P}{\partial t} &=&  \gamma \frac{\partial}{\partial v} {\left[ (v-\theta)P \right]} 
+ \frac{1}{2}\frac{\partial}{\partial x}(vP)+ \rho\kappa\frac{\partial^2}{\partial x\partial v}(vP)+\frac{1}{2}\frac{\partial^2}{\partial x^2}(vP) \nonumber \\ 
&+& \frac{\kappa^2}{2}\frac{\partial^2}{\partial v^2}(vP),
\end{eqnarray}
that has to be solved for the following boundary condition $P(x,v,0\mid v_i)=\delta(x)\delta(v-v_i)$.

A Fourier transform in $x$ followed by a Laplace transform in $v$ leads to a  partial differential equation of the first degree that can be solved by the method of characteristics \cite{zwillinger}: 
\begin{eqnarray}
\label{eqdiff}
&&\left\{\frac{\partial}{\partial t}+\left[\Gamma p_v+\frac{\kappa^2}{2} p_v^2 -\frac{(p_x^2-ip_x)}{2} \right] \frac{\partial}{\partial p_v}\right\}Q= -\gamma\,p_v\,Q,
\end{eqnarray}
where $\Gamma\equiv i \rho\kappa p_x + \gamma$ and the boundary condition is $Q(p_x,p_v,0\mid v_i)=e^{-p_v v_i}$.

The unconditional distribution of log-returns can be obtained by inverting the Laplace and Fourier transforms and integrating first over  the volatility $v$ and then over the initial volatility $v_i$:
\begin{equation}
\label{eq:inversion2}
P_t(x)= \int_0^\infty dv_i\;P^*_v(v_i)\int_0^{\infty} dv \int_{-\infty}^{+\infty}\frac{dp_x}{2\pi}\;e^{ip_xx}\,Q(p_x,0,t\mid v_i)\nonumber,
\end{equation}  
where $P^*_v$ is the stationary solution for the volatility distribution (see \cite{yakovenko} for details).

Integrating (\ref{eq:inversion2}) we finally get:
\begin{equation}
\label{px4}
P_t(x)=\int_{-\infty}^{+\infty}\frac{dp_x}{2\pi}\;e^{ip_xx}\exp\left[\alpha\left(\frac{\Gamma t}{2} - \ln\cosh\left(\frac{\Omega t}{2}\right) - \Phi_t(p_x)\right)\right],
\end{equation}
where
\begin{eqnarray}
\label{eq_Omega}
\Omega&=&\sqrt{\Gamma^2+\kappa^2\left(p_x^2-ip_x\right)}  \\
\Phi_t(p_x)&=&\ln\left[1+\frac{\Omega^2-\Gamma^2+2\gamma\Gamma}{2\gamma\Omega}
\tanh\left(\frac{\Omega t}{2}\right)\right]. 
\end{eqnarray}

\section{Volatility Autocorrelation and Leverage Functions}

The formal integral of (\ref{eq_volprocess}) is given by:
\begin{equation}
\label{eq_volprocess_int}
v_t= \left( v_0 - \theta \right) e^{-\gamma_1 t} + \theta + \kappa\int_0^{t}dW_1(u)\,e^{-\gamma(t-u)}\sqrt{v_u}. 
\end{equation} 
A simple calculation gives the stationary autocorrelation function: 
\begin{eqnarray}
\label{AC_function}
C(\tau\mid\gamma,\theta,\kappa)\equiv  \lim_{t\rightarrow\infty}\frac{\langle v_t v_{t+\tau}\rangle-\langle v_t\rangle\langle v_{t+\tau}\rangle }{\theta^2}=\frac{e^{-\gamma\tau}}{\alpha}.
\end{eqnarray}

The leverage function is \cite{perello}:
\begin{equation}
\label{leverage}
L(\tau\mid\gamma,\theta,\kappa,\rho)\equiv \lim_{t\rightarrow\infty} \frac{\langle dx_t \left(dx_{t+\tau}\right)^2\rangle}{{\langle\left(dx_{t+\tau}\right)^2\rangle}^2} = \rho \kappa\, H(\tau)\, G(\tau)\,e^{-\gamma\tau},
\end{equation}
where $dx_t$ is given by (\ref{eq_detrended}), $H(\tau)$ is the Heaviside step function and: 
\begin{equation}
G(\tau)=\frac{\left\langle v_t \exp\left[\frac{\kappa}{2}\int_{t}^{t+\tau} dW_1(u)\, v_u^{-\frac{1}{2}}  \right]\right\rangle}{{\langle v_t \rangle}^2}.
\label{eq_G}
\end{equation} 
To simplify the numerical calculations  we employ throughout this paper the zeroth order appoximation $G(\tau)\approx G(0)= \theta^{-1}$. The approximation error increases with the time lag $\tau$ but is not critical to our conclusions.

In the next section we simultaneously fit  (\ref{AC_function}), (\ref{leverage}) and  the model (\ref{px4}) to fluctuations of the BOVESPA  index at a  wide range of time scales.

\section{Fitting IBOVESPA data}

\begin{figure}
\begin{center}
\epsfig{file=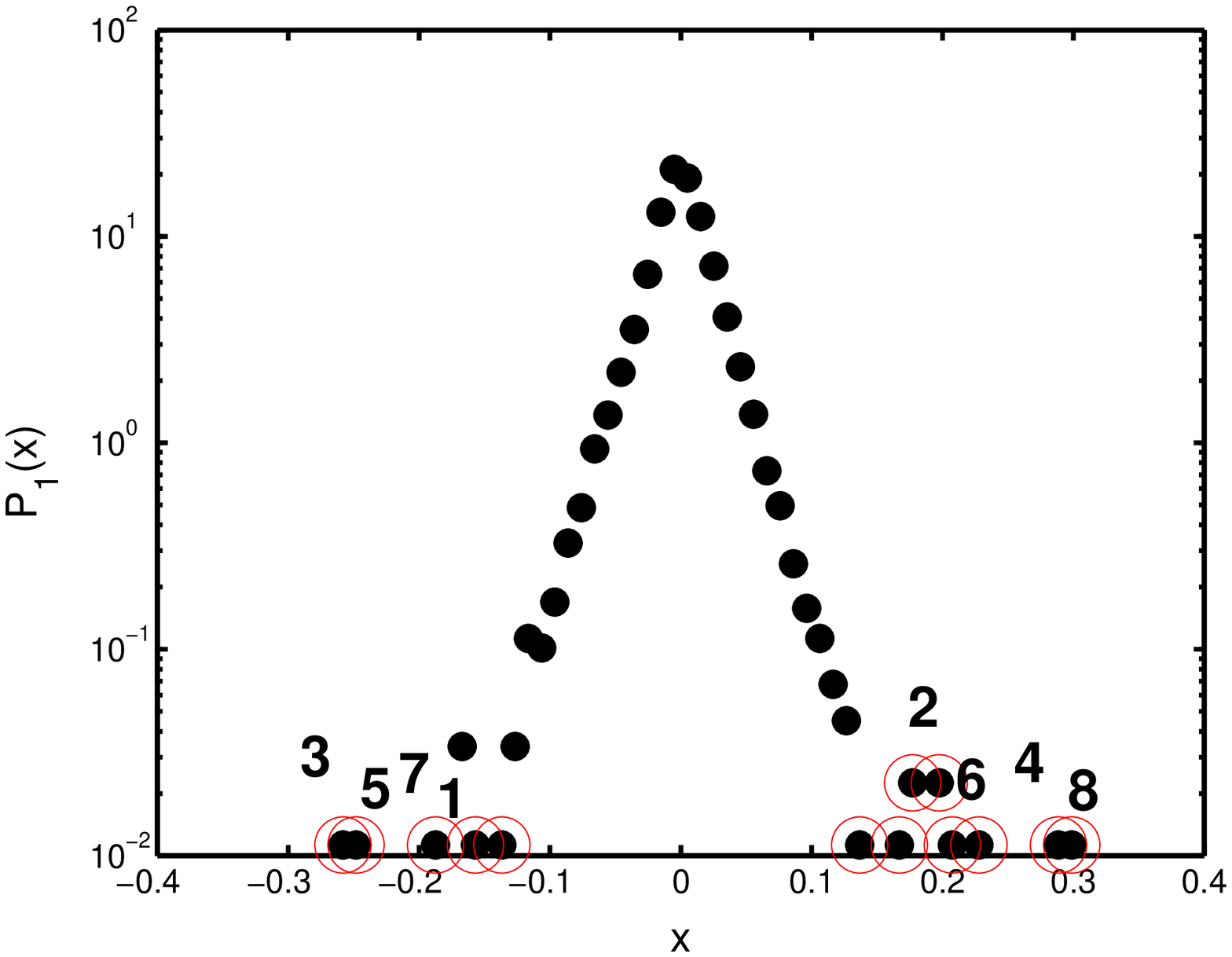,width=70mm}\epsfig{file=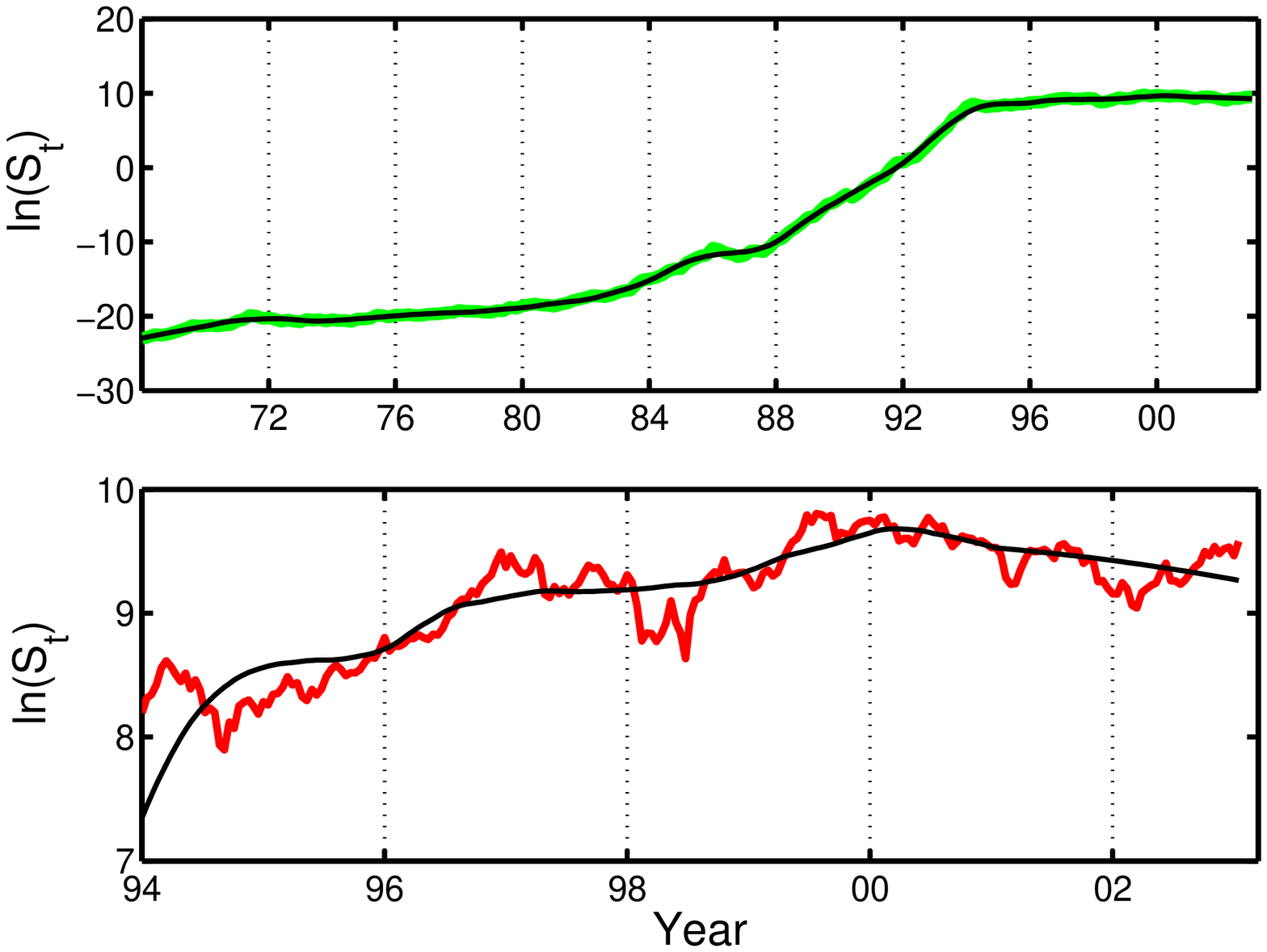,width=70mm}
 \caption {Left: Large deviations in daily log-returns with identifiable exogenous causes:  New York Stock Market Crash (1);  Economic Plans (2-4); Presidential crises (5), Recovery from Mexican, Asian  and Russian crises (6); Asian and Russian crises (7) and  Currency crises (8). Right: (Top) Macroeconomic trend in the log-prices. (Bottom) Trend and fluctuations in the last ten years.} \label{data_prepare}
\end{center}
\end{figure}

\subsection{The data}
Three data sets were used. IB1 consists of daily data from IBOVESPA inception on January, 1968 to December, 2002. IB1 was adjusted to take into account eleven  divisions by 10 introduced in the period for disclosure purposes \cite{ibovespa}. IB2 consists of daily data from January, 2001 to August, 2003. IB3 consists of high-frequency data from March 3, 2001 to February 14, 2003 and from June 6, 2003 to August 26, 2003.

\subsection{Trimming}
Large deviations are explained both by endogenous dynamics and by exogenous shocks. An adequate microscopic model for the price formation process  has to explain spontaneous large movements and the system response to external shocks. However, here we only  intend to model the typical behavior at mesoscopic scales (price dynamics).   In emerging markets, large movements  with exogenous identifiable causes are frequent. In order to obtain a reliable fitting  for the typical fluctuations we expunged from the data set large deviations connected to major structural changes, in Figure 1 we identify those externalities. For a brief historical  account of recent Brazilian economy see \cite{costa}.

\subsection{Filtering the Macroeconomic Dynamics}
In Equation \ref{eq_detrended} we extract the macroeconomic drift $\mu_t$ to focus on price fluctuations. This drift at long time scales reflects the effect of inflation, economic growth, business cycles and the riskless interest rate. To extract $\mu_t$ we have employed a low-pass Savitzky-Golay smoothing filter of degree two \cite{nr} with averaging over a four years moving window (\textit{i.e.} one Brazilian presidential term). In Figure 1 we show $\int_0^t d\tau\,\mu_{\tau}$ for the whole data set (top) and the fluctuations around the trend in the last ten years (bottom). This choice of smoothing technique is  heuristic and may introduce artifacts at  time scales of lengths comparable to the moving windows (4 years in our case). We, therefore, restrict our analysis to a maximum period of 100 days and leave a thorough analysis of this aspect to another occasion.  

\subsection{Fitting Parameters}

\begin{figure}
\begin{center}
\epsfig{file=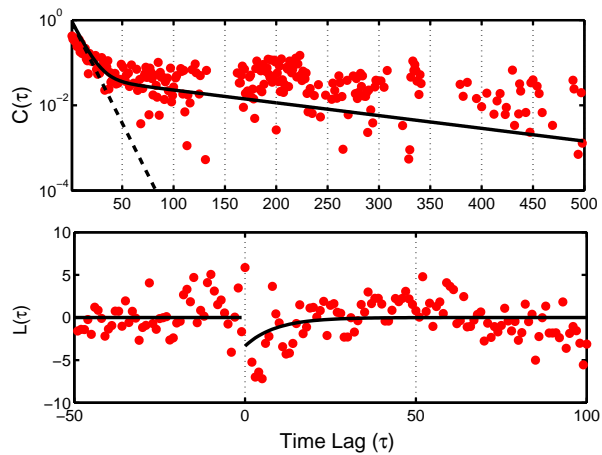,width=80mm}
 \caption {Top: Volatility autocorrelation: IB1 (circles), Heston model (dashed) and two time scales (solid line).  Bottom: Leverage function: IB1 (circles) and approximate Heston model (see text, solid line).} \label{data_LEVACF}
\end{center}
\end{figure}

After filtering the trend we have to fit four parameters: the long term mean volatility $\theta$, the relaxation time for mean reversion $1/\gamma$, the volatility fluctuation scale $\kappa$ and the correlation between price and volatility $\rho$. It became apparent in \cite{silva}  that a simple least squares fit that  takes into account only (\ref{px4}) yields parameters that are not uniquely defined. The task of finding parameters fitting  all the main stilyzed facts simultaneously is not trivial. In this paper we do not intend to focus on parameter estimation, hence we adopt a heuristic procedure and do not provide rigorous error bars. These questions will be addressed somewhere else. We first estimate  the long term mean volatility  $\theta$ directly from the daily log-returns as:
\begin{equation}
	\hat{\theta}=\frac{1}{N-1}\sum_{j=1}^N {x^{(1)}_j}^2,
	\label{var}
\end{equation}
where $x^{(1)}$ stands for daily detrended log-returns. 

A proper fit must  be also consistent  with the non-vanishing volatilities  constraint, $\frac{d}{2}>\frac{2\gamma \theta}{\kappa^2}\geq 1$. We, therefore, minimize a constraint free cost function and adjust $\kappa$ to attain consistency with the constraint.

\begin{figure}
\begin{center}
\epsfig{file=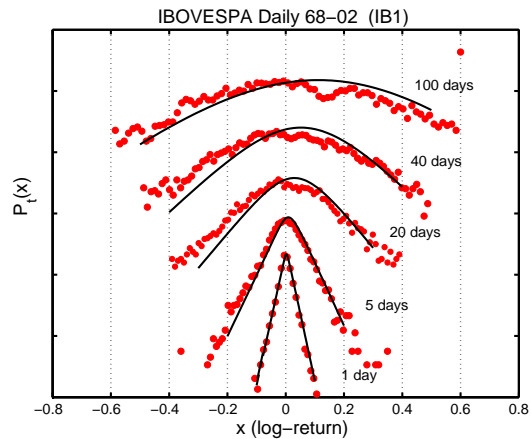,width=70mm}
 \caption{Probability densities of IBOVESPA returns from $1$ to $100$ days for the dataset IB1. The vertical scale is logarithmic and is multiplied by a constant for better visualization. } \label{IB1}
 \label{PDF}
\end{center}
\end{figure}

The following  cost function is employed:
\begin{eqnarray}
\fl	E(\gamma,\kappa,\rho)&\equiv&  \frac{1}{I}\sum_{i=1}^{I} \left[1-\frac{L^{E}(\tau_i)}{L(\tau_i\mid \gamma,\hat{\theta},\kappa,\rho)} \right]^2 + \frac{1}{L}\sum_{l=1}^{L} \left[1-\frac{C^{E}(\tau_l)}{C(\tau_l \mid \gamma,\hat{\theta},\kappa)} \right]^2.
	\label{cost}
\end{eqnarray}
 The empirical  leverage function is \cite{perello}: 
\begin{equation}
L^{E}(\tau) = \frac{\frac{1}{M}\sum_{t=1}^M  x^{(1)}_t \,\left( x^{(1)}_{t+\tau}\right)^2}
{\hat{\theta}^2},
\end{equation}
where $x^{(1)}_t=\ln\left(\frac{S_t}{S_{t-1}}\right) - \mu_t$ and the empirical autocorrelation function is:
\begin{equation}
C^{E}(\tau) = \frac{\frac{1}{M}\sum_{t=1}^M  \left( x^{(1)}_{t}\right)^2 \,\left( x^{(1)}_{t+\tau}\right)^2}
{\hat{\theta}^2}.
\end{equation}
We then employ the parameters obtained above to fit the probability density of returns and adjust the  parameter $\rho$ to the empirical data minimizing the mean squared error. We will discuss a more systematic approach to parameter estimation elsewhere as  our main aim here is restricted to showing the description capability of the Heston model. 

The parameters found are in the following table:
\begin{center}
\begin{tabular}{|l|l|l|}\hline\hline
\emph{parameter} & \emph{IB1}&\emph{IB2-IB3}\\\hline
$\theta$ & $7.8\times 10^{-4}$ $days^{-1}$ & $5.2\times 10^{-4}$ $days^{-1}$\\\hline
$1/\gamma$ & $9.0$ $days$ & $5.8$ $days$  \\\hline
$\rho$ & $-0.20$ & $-0.15$ \\\hline
$\kappa$ & $1.3 \times 10^{-2}$ $days^{-1}$ & $1.1 \times 10^{-2}$ $days^{-1}$   \\\hline
$d$ & 2.03& 2.98\\\hline\hline
\end{tabular}
\end{center}
Figure \ref{data_LEVACF} shows IB1 data and curve fits. It is clear that, under the vanishing volatility constraint, a single relaxation time is not simultaneously consistent with  leverage and autocorrelation functions. The best fit for the autocorrelation is shown as a dashed line. We deal with this inconsistency in the final sections of this paper by extending the Heston model to include two relaxation times.

\subsection{Intraday Fluctuations}

\begin{figure}
\begin{center}
\epsfig{file=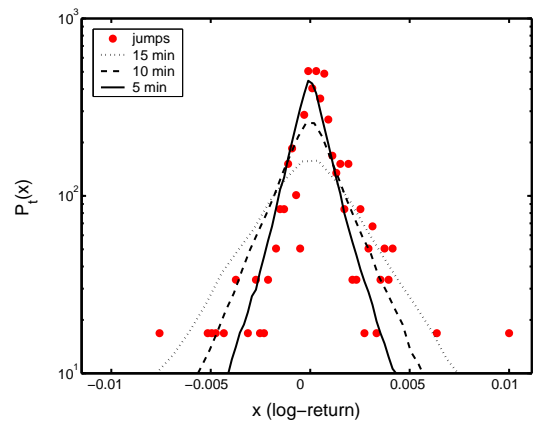,width=70mm}
 \caption{Probability densities of  returns at $5,10$ and $15$ minutes,  overnight and lunch time jumps (dataset IB3).} \label{jumps}
\end{center}
\end{figure}

\begin{figure}
\begin{center}
\epsfig{file=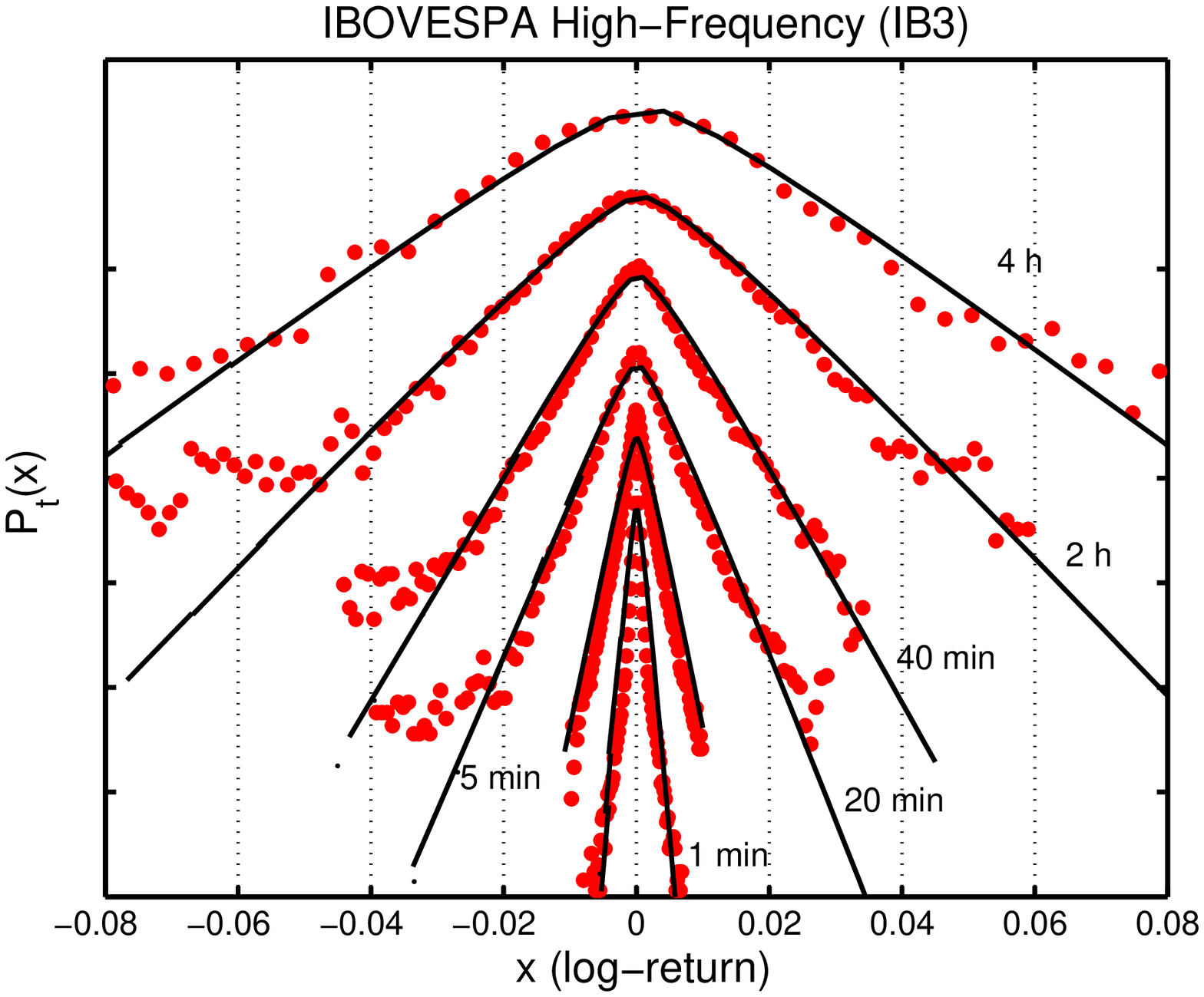,width=70mm}\epsfig{file=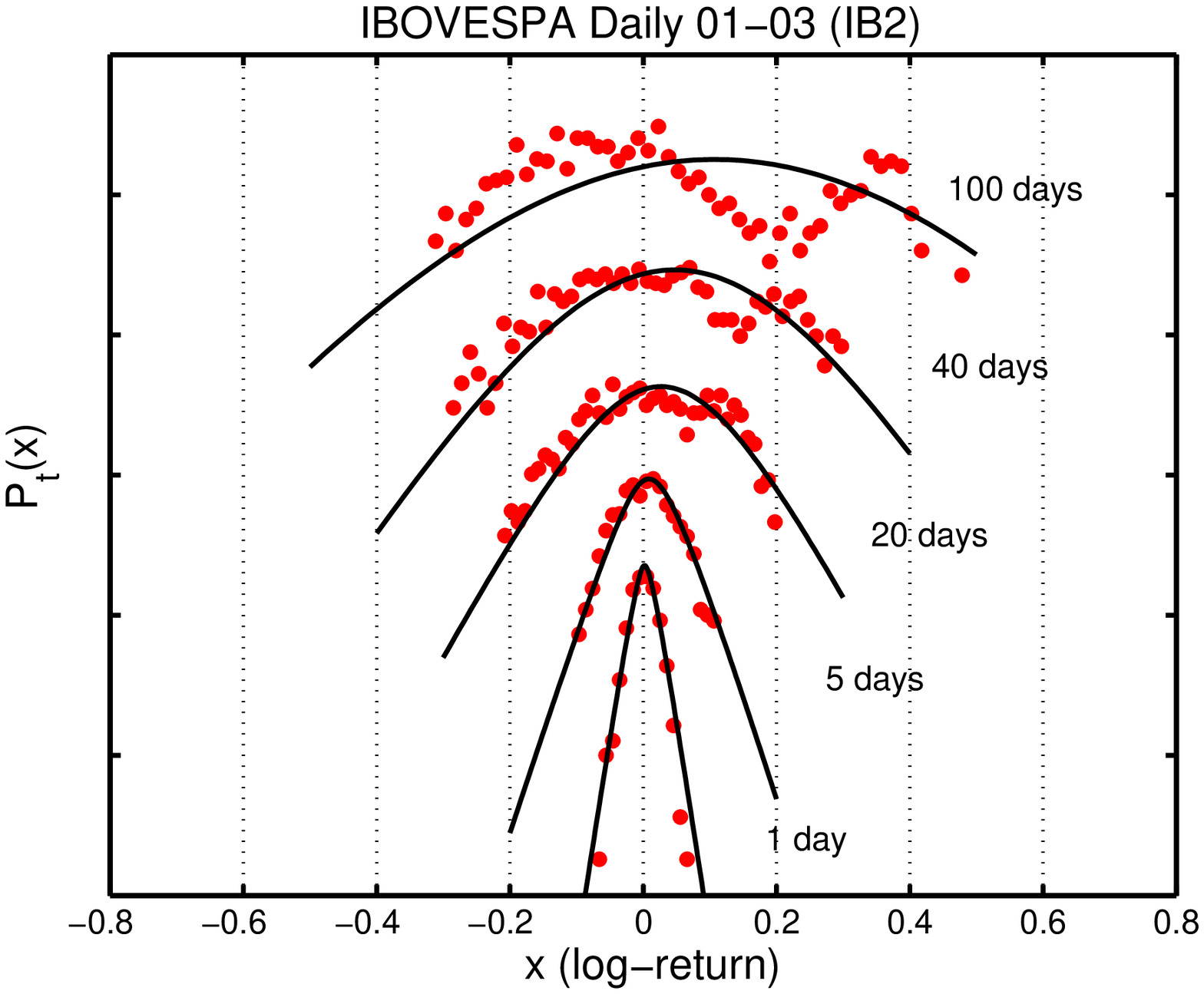,width=70mm}
 \caption{Left: Probability densities for intraday returns from $1$ minute to $4$ hours. Circles are empirical data from IB3 while lines correspond to the theory. Right:  Returns from $1$ to $100$ days. Circles are empirical data from IB2. A single set of parameters (see Table) is used through all time scales considered. The vertical scale is logarithmic and is multiplied by a constant for better visualization. }
 \label{IB2}
\end{center}
\end{figure}

At first glance, it is not clear  whether intraday and daily returns can be described by the same stochastic dynamics. Even less clear is whether  aggregation from intraday to daily returns can be described by the same parameters. To verify this latter possibility we have to transform units by determining the effective duration in minutes of a business day  $T_{eff}$. The dimensional parameters are, therefore, $\theta^{(ID)}=\theta/T_{eff}$, $\gamma^{(ID)}=\gamma/T_{eff}$ and $\kappa^{(ID)}=\kappa/T_{eff}$. In order to avoid the effects of non-stationarity we also use data from the same period (IB2 and IB3). From IB2 and IB3  we found  the effective duration of a day to be about $T_{eff}=540$ minutes.  The S\~ao Paulo Stock Market opens daily at 11 a.m. and closes at 7:30 p.m. local time, resulting in exactly $510$ minutes. The effective duration infered from data  exceeds, therefore, in $30$ minutes  the real duration of a normal business day, an amount that is  explained by the effect of overnight and lunchtime jumps.  In Figure 4  we show a histogram of jumps compared to intraday returns at $5,10$ and $15$ minutes. Considering both overnight and lunchtime jumps, between $10$ to $30$ minutes must be added to a normal business day, just the right amount to make $T_{eff}$ compatible to the hypotheses of a single set of parameters describing  time scales from minutes to months. In Figure \ref{IB2} we show observations and data for IB2 and IB3 with parameters given by the table above. These findings are suported by similar results presented in \cite{silva2} for high-frequency returns of individual stocks traded in NASDAQ and NYSE.
 
\begin{figure}
\begin{center}
\epsfig{file=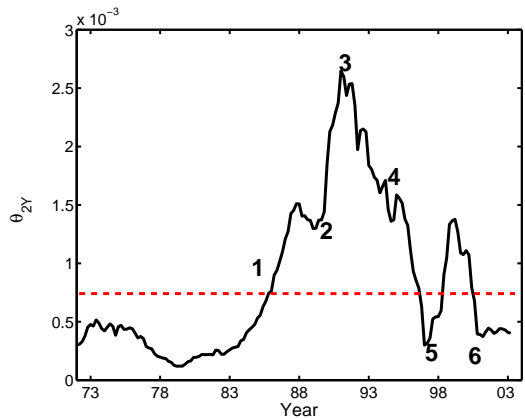,width=70mm}
 \caption{ Historical perspective for the non-stationarity of the mean volatility  calculated using the dataset IB1 and a two years wide moving window. The numbers identify the main political events as follows: military dictatorship ends after two decades (1), first presidential election (2), presidential impeachment (3), second election (4), reelection (5), third election (6).} \label{histvol}
\end{center}
\end{figure}

The datasets IB1 and IB3 yield $T_{eff}=600$ minutes, the $90$ minutes excess is explained by the well known non-stationarity of the parameters revealed in the table above, that shows that the mean variance in the $35$ years period  (IB1, $\theta=7.8\times 10^{-4}$ $days^{-1}$) is  higher than the mean variance observed in the last two years (IB2, $\theta= 5.2\times 10^{-4}$ $days^{-1}$). It is interesting to stress that  in an emerging market like the Brazilian structural changes related to political events are frequent. In Figure \ref{histvol} we try to put the the macroeconomic non-stationarity of the  mean variance $\hat{\theta}$ in a historical perspective, identifying some important political events. The robust statistical behavior here reported are particularly surprising when one considers the extremely abrupt structural changes that take place in an emerging market.

\section{Multiple Time Scales}

It is clear in Figure \ref{data_LEVACF} that the Heston model is not capable of describing the autocorrelation function. To rectify this inconsistency we propose the inclusion of a second, slower, time scale into the dynamical model along the lines of \cite{bouchaud}. 
We assume a stochastic dynamical model where the stochastic volatility reverts
to a second stochastic volatility with much longer relaxation time $\gamma_1\gg\gamma_2$. The new model reads:
\begin{eqnarray}
\label{eq_heston_2scales}
dS &=& S(t)\mu dt\;+\;S(t)\sqrt{v(t)}\;dW_0(t)\\
dv &=& -\gamma_1\left[v(t)-\theta\right]dt\;+\;\kappa_1\sqrt{v(t)}\;dW_1(t),\nonumber\\ 
d\theta &=& -\gamma_2\left[\theta(t)-\theta_0\right]dt\;+\;\kappa_2\sqrt{\theta(t)}\;dW_2(t),\nonumber
\end{eqnarray}
where $\gamma_2/\gamma_1 << 1$ e $dW_j$ are Wiener processes defined as:  
\begin{eqnarray}
\label{eq_Wiener_2scales}
\langle dW_j(t) \rangle &=& 0,\\
\langle dW_j(t)\, dW_k(\tilde{t}) \rangle &=& C_{jk}\delta(t-\tilde{t})\;dt\nonumber,
\end{eqnarray}
where $C$ is a correlation matrix with $C_{jj}=1$, $C_{12}=\rho$ and $C_{13}=C_{23}=0$.

In this model the autocorrelation function acquires the following form:
\begin{eqnarray}
\label{AC_function_2scales}
C(\tau)= \frac{e^{-\gamma_1\tau}}{\alpha_1}+\frac{e^{-\gamma_2\tau}}{\alpha_2},
\end{eqnarray}
with $\alpha_2=\frac{2\gamma_2\bar{\theta}}{\kappa_2^2}$, where $\bar{\theta}$ stands for the average of $\theta$ given $\theta_0$. In Figure \ref{data_LEVACF} we show a fit of this autocorrelation function to the data, the new relaxation time is $\gamma_2^{-1}=144.9$ days and $\kappa_2=1.0\times 10^{-4}$ $\mbox{days}^{-1}$. It is possible to solve the Fokker-Planck equation for this extended model in the limit where $\alpha_2\gg\alpha_1$. We will discuss this model in detail elsewhere \cite{vicente}.

\section{Conclusions}
\label{conclusion}
We showed that the Heston model can reproduce the diffusion process of a 35 years long time series of IBOVESPA returns at a wide range of time scales from days to months. We also show that the Heston model can explain the aggregation of returns from minutes to months with a single set o parameters that change in the macroeconomic scales. However, the Heston model, with a single relaxation time, is not capable of explaining the behavior of the autocorrelation function. We, therefore, proposed an extended version of the model with the addition of a slow stochastic dynamics that yields an autocorrelation function consistent with the observations. 

It is surprising  that a single non-trivial stochastic model may be capable of explaining the long-term statistical behavior of both  developed and emerging markets, despite the known instability and high susceptibility to externalities of the latter. We believe that this robust non-trivial statistical behavior is evidence for more basic mechanisms acting in the market microstructure. Perhaps the search for underlying common mechanisms (or laws) that can explain empirical data  is the main contribution of Physics  to  Economics. This contribution might be particularly useful to the field of Econometrics in which a common view  is that a theory built from data `should be evaluated in terms of the quality of the decisions that are made based on the theory' \cite{granger}. Clearly, these two approaches should not be considered as mutually exclusive.

\ack
We thank Victor Yakovenko and his collaborators for discussions and for providing useful MATLAB codes. We also wish to thank the S\~ao Paulo Stock Exchange (BOVESPA) for gently providing high-frequency data. This work was partially (RV,VBPL) supported by FAPESP.


\end{document}